\documentclass[10pt,conference,letterpaper,twoside,final]{IEEEtran}

\usepackage{times}
\usepackage{url}

\usepackage{amssymb}
\usepackage{amstext}
\usepackage{amsmath}

\usepackage{setspace}

\usepackage{mdwlist}

\usepackage{subfigure}
\usepackage[pdftex]{graphicx}
\DeclareGraphicsExtensions{.pdf,.gif}

\usepackage{setspace}

\usepackage{todonotes}

\title{Profit-Aware Server Allocation\\for Green Internet Services}

\begin{document}

\author{
\IEEEauthorblockN{
Michele Mazzucco\IEEEauthorrefmark{1}\IEEEauthorrefmark{3},
Dmytro Dyachuk\IEEEauthorrefmark{2} and
Marios Dikaiakos\IEEEauthorrefmark{1}
}
\IEEEauthorblockA{\IEEEauthorrefmark{1}University of Cyprus, Cyprus}
\IEEEauthorblockA{\IEEEauthorrefmark{2}University of Saskatchewan, Canada}
\IEEEauthorblockA{\IEEEauthorrefmark{3}University of Tartu, Estonia}
}

\maketitle

{ \abstract
A server farm is examined, where a number of servers are used to offer a service
to impatient customers. Every completed request generates a certain amount of
profit, running servers consume electricity for power and cooling, while waiting
customers might leave the system before receiving service if they experience
excessive delays. A dynamic allocation policy aiming at satisfying the
conflicting goals of maximizing the quality of users' experience while
minimizing the cost for the provider is introduced and evaluated. The results of several experiments
are described, showing that the proposed scheme performs well under different
traffic conditions. \endabstract }

\section{Introduction}
\label{sec:introduction}

In recent years large investments have been made to 
build data centers, or {\it server farms}, purpose-built facilities providing
storage and computing services within and across organizational boundaries. 
A typical server farm may contain thousands of servers, which require large
amounts of power to operate and to keep cool, not
to mention the hidden costs associated with data centers' carbon footprint and water consumption for cooling purposes~\cite{hamilton:2009a}. 
On the other hand, the increasing use of the Internet as a provider of services
and a major information media have changed significantly, especially over the
last ten years. Expectations in terms of performance
and responsiveness have markedly grown. 
For example, Google reports that an extra 0.5 seconds in search page generation entails degraded user satisfaction,
with a consequent 20\% traffic drop, while trimming the page size of Google Maps by 30\% resulted in a 30\% traffic increase~\cite{linder:2006a},~\cite{shankland:2008}.
Hence, the development of `green' data centers, {\it
i.e.}, data centers that are energy efficient, is a challenging problem for
service providers as they operate under stringent performance requirements. 
Consequently, it is very important to devise strategies aiming at reducing the power
consumption while maintaining acceptable levels of performance.

Unfortunately, despite considerable effort in designing servers whose power consumption is
proportional to their utilization~\cite{barroso:2007}, the reality is that the amount of power
consumed by an idle server is about 65\% of its peak
consumption~\cite{greenberg:2009}, as
existing hardware components offer only limited controls for trading power for
performance. Thus, the only way to significantly reduce
data centers' power consumption is to improve the server farm's utilization,
{\it e.g.}, by tearing down servers in excess. Therefore, we propose and
evaluate a strategy that aims at maximizing the overall performance while minimizing the number of required servers.
Under suitable assumptions about the nature of user demand, it is possible
to explicitly evaluate the effect of a particular server allocation on the
achievable revenue. 
Hence, we derive a numerical algorithm for computing the optimal number of
servers required for handling a certain user demand. The model
considers limited user patience time and the fact that servers energy
consumption depends on servers' utilization. The computational costs of the decision making
are extremely low and the algorithm can be used on-line as a part of a
dynamic allocation policy.


The problem of reducing energy consumption of server farms can be
approached from different angles. Improving energy efficiency for servers
by means of dynamic scaling of the CPU frequency has been addressed in
several papers, {\it e.g.}, \cite{Elnozahy:2002,
Sharma:2003}. An alternative solution consists of switching off servers in
excess. The most closely related work can perhaps be found in~\cite{chen:2005}
and~\cite{mazzucco:2010a}. The former discusses a queuing model for controlling
the energy consumption of service provisioning systems subject to Service Level
Agreements. However, while Chen et al. take into account the cost for smaller
mean time between failures (MTBF) when powering up/down some servers, the cost function they
propose does not consider the time and energy wasted during state changes, nor
the cost for failing to meet the promised quality requirements. Hence, the taken
decisions could be
either too performance oriented or too energy-efficiency oriented. 
\cite{mazzucco:2010a}, instead, discusses a problem similar to that we attack in
this paper. However, in that paper the authors assume that clients have no
patience, while they do not consider the fact that servers consume energy
without producing any revenue during system reconfigurations. 
Finally, since
running too many servers increases the electricity consumption while having 
too few servers switched on requires running those servers' CPUs at higher 
frequencies, some hybrid approaches have been proposed, {\it
e.g.},~\cite{Elnozahy:2002}.

The rest of the paper is structured in the following way. In the next
section we introduce the system model. Section~\ref{sec:analysis} contains
the mathematical analysis. The model for power consumption
estimation is discussed in Section~\ref{sec:power}, while the resulting policy
is introduced in Section~\ref{sec:policies}. A number of
experiments are presented in Section~\ref{sec:experiments}. Finally, we
conclude the paper in Section~\ref{sec:conclusions} with a summary and
some remarks.

\section{The Model}
\label{sec:model}

A server farm is a collection of servers 
interconnected by high-speed, switched LANs that hosts content and runs 
 applications (or services) accessed over the Internet. 
%
%
%
In this paper we focus on server farms designed according to the dedicated
architecture, where a web application is hosted on a set of
physical servers (see, for example, \cite{mazzucco:2010b}) and
the provider can change the number of servers
allocated to run each service in order to react to traffic changes. Once a decision about how to partition
the available servers has been made, it is possible to treat
each subsystem ({\it i.e.}, service) in isolation of each other. Therefore, in
the following we tackle the problem of maximizing the revenue of a single
subsystem (service) only.

Now, assume that $q$ (identical) servers have been allocated to queue $i$.
Among those $q$ servers, $h$ are running and capable of serving incoming user demand ({\it jobs}, from now on), while the remaining $(q - h)$ are switched off in order to save energy. In this context, the term `switched off' means that the server can
not perform any useful work and it does not consume any electricity.
A maximum of $m$ jobs can be processed in parallel on each server without
significant interference -- this limit being imposed by the number of available
threads or processes. This is modelled by assuming that there are $m$ parallel
servers on each physical machine or core, and thus a total of $S = qm$ servers are available, while $n = hm$ are running. If $n$ jobs are currently in execution, further requests are
temporarily parked into an external first-in-first-out (FIFO) queue whose size is
assumed to be infinite. 
If, once a server has finished processing a request, the
queue is empty, the server begins to idle ({\it i.e.}, it consumes energy
without generating any profit). Otherwise, it removes the leading job from the
waiting queue and starts processing it.
Every processed request generates some profit. For example, it can
be a profit coming from advertisements or from sales (in case of online
merchants, such as Amazon). While in the first case advertising agencies usually
pay for each impression ({\it i.e.}, display of an add), in the second case the
profit from each request can be estimated as follows. Suppose that
every 10,000 views generate 10\$ of revenue from sales. Hence, we can state
that on average each request brings 0.01 cents. It is worth stressing that
companies like eBay employ more complex revenue models. However, using simple
transformations such as dividing gross income over the number of requests, one
can easily estimate the average profit brought by every request.

\begin{figure}[ht!]
\centering
\includegraphics[width=0.4\textwidth]{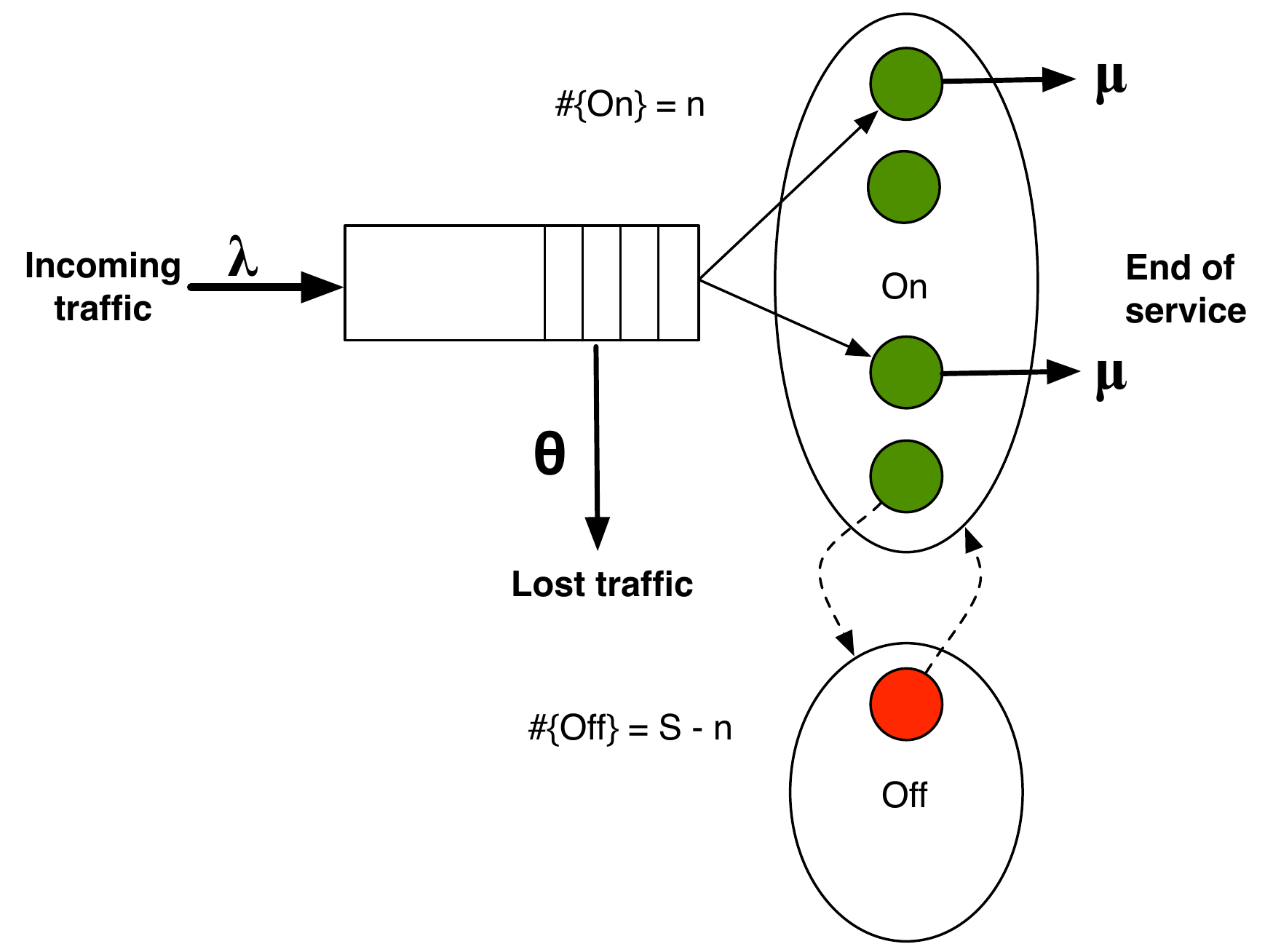}
\caption{System model. Jobs whose average job size is $1 / \mu$ enter the
system at rate $\lambda$ and abandon the system at rate $\theta$ while
waiting.}
\label{fig:model}
\end{figure}

Since running servers consume electricity, which costs $r$\$ per kWh, the
provider dynamically decides how many servers to run by means of a `resource
allocation' policy. The objective is to find the optimal number of servers, $n$,
that should be switched on in order to optimize the provider's profit. The
extreme values, $n=0$ and $n=S$, correspond to switch respectively off, or on,
all available servers. Given that the amount of running servers should change in
response to changes in the user demand, the problem is how to estimate the best
$n$. We assume that data is widely replicated, hence switching some servers off
does not affect service availability. However, since lost jobs do not
generate any revenue, the provider should ensure that the time users wait for their
requests to be served does not exceed their patience. Otherwise the clients will
start aborting their requests, {\it e.g.}, by clicking a Stop button in their
browsers, see Figure~\ref{fig:model}.

During the intervals between consecutive policy invocations,
the number of running servers remains constant.
Those intervals, which will be referred to as `observation windows', are used 
by the controlling software to collect the traffic statistics used by the
allocation policy at the next decision epoch.


While different metrics can be used to measure the performance of a computing 
system, as far as the service provider is concerned, the performance of the 
server farm is measured by the average revenue, $R$, earned per unit time. 
That value can be estimated as

\begin{equation}
R = c T - rP \mbox{,}
\label{eq:R}
\end{equation}
 
\noindent where $c$ is the income generated by each completed job, $T$ is the system's
throughput, and $P$ is the total average power consumed by powered up
servers. $c$ is a parameter of the model, while the formula for computing $T$ and $P$ are discussed in Sections~\ref{sec:analysis} and~\ref{sec:power}.
For the following it will be convenient to indicate explicitly the dependency
of Equation~\eqref{eq:R} on the parameter $n$ by introducing the notation

\begin{equation}
R = r(n) \mbox{,}
\end{equation}

\noindent where $r(n)$ stands for the right-hand side of~\eqref{eq:R}.

It is worth noting that although we do not make any assumption about
the relative magnitudes of charge and cost parameters, the
most challenging case is when they are close to each other. If
the charge for executing a job is much higher than the provisioning cost, one could guarantee a positive, but not optimal, revenue by switching on all servers, regardless of the load. On the other hand, if the charge is
smaller than the cost, than it would be better to switch all
servers off.

\section{Analysis}
\label{sec:analysis}
\begin{figure*}[th!]
\centering
\includegraphics[width=0.85\textwidth]{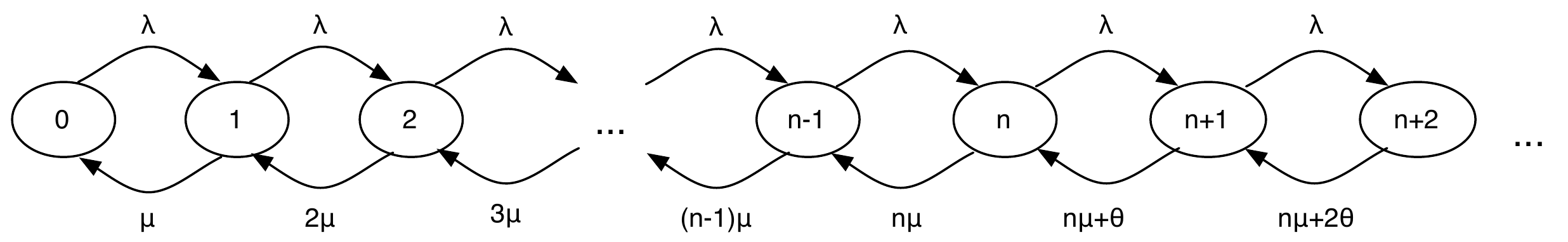}
\caption{State transition diagram.}
\label{fig:markov_chain}
\vspace{-4mm}
\end{figure*}

Suppose that $n$ servers have
been allocated to serve incoming traffic. Jobs enter the system according to an
independent Poisson process with rate $\lambda$. Operating servers accept one job at a time, with
the service times, or `job size', being exponentially distributed with mean $1
/ \mu$. Hence, one might try to model the system as an $M/M/n$
queue (see~\cite{mitrani:1998} for more details) with the offered load being $\rho =
\lambda /\mu$
%
%
and with the
stability condition being $\rho < n$. Unfortunately, this
system, also called Erlang-C (or delay), does not acknowledge abandonment; thus it either distorts
or completely fails to provide important information. 

In this work we do allow customers abandonment by assuming that a time-out policy
is in operation: if a job entering the system does not acquire a server  before
its time-out period expires, the job is terminated and leaves the system without
generating any revenue. This can be used to model HTTP time-outs, as well as
impatient customers. Impatient customers are of particular importance, as
\cite{mcgovern:2008} reports that 75\% of people would not go back to a web site
that took more than 4 seconds to load. HTTP time-outs are in practice of fixed
length, while users' patience is not. For the purposes of analytical
tractability, we assume that both the user's patience and HTTP time-outs are
i.i.d random variables distributed exponentially with mean $1/\theta$, with
$\theta$ being referred to as the \emph{abandonment rate}. The extreme values,
$\theta = 0$ and $\theta = \infty$, correspond to jobs with no, or infinite
patience. Hence, the appropriate queueing models would be $M/M/n/n$ and $M/M/n$
respectively~\cite{mitrani:1998}. Also, we assume that the patience variables are
independent of all other model elements, namely arrival and service rates.

The first step to computing Equation~\eqref{eq:R} is the steady
state analysis of the Markov chain associated with the model sketched in
Figure~\ref{fig:model}, see Figure~\ref{fig:markov_chain}. 
A tractable way to model this system is the
$M/M/n+M$ queue, also known as Erlang-A (where the `A' stands for `Abandonment')~\cite{palm:1957}. 
The main difference between the  Erlang-A and Erlang-C models is that 
the queue never grows unbound in the Erlang-A model, as jobs are allowed to
leave. Moreover, jobs abandonment reduces workload only when needed, {\it i.e.}, when the load is high. 
The implication is that fewer servers are needed to guarantee the same level of
performance under Erlang-A, compared to the traditional Erlang-C.
This observation is very important because in this paper we focus on highly loaded (and
potentially overloaded) systems, as our goal is to switch off servers in
excess while serving as many customers as possible. 

Now, let $L_t$ be the
total number of jobs inside the system at time $t$, including both queueing
and executing requests. Then, $L=\{L_{t}: t \ge 0 \}$ is a state-dependent
{\it Birth-and-Death} process~\cite{mitrani:1998}. 
The instantaneous transition rate from state $j$ to state $(j +
1)$ is equal to the arrival rate, $\lambda_{j} =
\lambda$, $j = 0, 1, \ldots$.
The conditional departure rate, {\it i.e.}, the transition rate from state $j$
to state $(j-1)$, depends instead on the number of operating servers as well as
on the number of jobs present. Hence, we should distinguish between two cases:

Case 1: $j \le n$. The system behaves like an $M/M/\infty$ queue: all jobs in
the system are being served without queueing, jobs leave the system at rate
$\mu_{j} = j\mu$, and $(n - j)$ servers are idle.

Case 2: $j > n$. All servers are busy and $(j - n)$ jobs are queueing. The
instantaneous completion rate does not depend on $j$ anymore, while the abandonment rate depends on the current number of jobs
in the queue. 

Hence, the balance equations can be expressed in
terms of $p_{0}$, {\it i.e.}, the probability of the system being empty

\begin{equation}
\label{eq:bal_eq}
p_{j} = \left\{ 
\begin{array}{ll}
\displaystyle{\frac{\rho^{j}}{j!}}p_{0} & \textrm{if $j \le n$}\\
\\
\displaystyle{\frac{\rho^{n}}{n!}} p_{0} \left[\prod_{k=n+1}^{j}
\frac{\lambda}{n\mu + \theta(k-n)} \right] &
\textrm{if $j > n$} 
\end{array} \right. \mbox{.}
\end{equation}

The only unknown probability is $p_{0}$.
Steady state for this process exists only if Equation~\eqref{eq:bal_eq} can be
normalized, {\it i.e.}, if $\sum_{j=0}^{\infty} p_{j} = 1$.
From the normalization condition and from Equation~\eqref{eq:bal_eq} we obtain

\begin{equation}
\label{eq_p_0}
p_{0} = \left[\sum_{j=0}^{n} \frac{\rho^{j}}{j!} + \frac{\rho^{n}}{n!}
\sum_{j=n+1}^{\infty} \prod_{k=n+1}^{j} \left( \frac{\lambda}{n\mu +
\theta(k-n)}\right) \right]^{-1} \mbox{.}
\end{equation}

Thus, steady state for this Birth-and-Death process always exists, as jobs in
excess eventually abandon, ensuring that the queue never grows unbound. To handle
the series in Equation~\eqref{eq_p_0} it is convenient to introduce the
function~\cite{palm:1957}

\begin{equation}
\label{eq_A}
g(x,y) = 
1 + \sum_{j=1}^{\infty}
\frac{y^{j}}{\displaystyle{\prod_{k=1}^{j} (x+k)}} \mbox{.}
\end{equation}

Thus, by means of Equation~\eqref{eq_A} and the formula for computing the
blocking probability in an Erlang-B system with $n$ trunks and traffic intensity
$\rho$, $B(n, \rho)$~\cite{mitrani:1998, hudousek:2003}, after some algebraic
manipulations we obtain

\begin{equation}
\label{eq_p_0_1}
p_{0} = \frac{n!}{\rho^{n}} \frac{B(n, \rho)}{1+B(n, \rho) \left[g
\displaystyle{\left(\frac{n\mu}{\theta}, \frac{\lambda}{\theta} \right)}
- 1\right]}
\mbox{.}
\end{equation}

Having computed the steady state probability of the number of jobs present, it is
possible to estimate the probability of abandonment, $P(Ab)$. Denote by
$P_{j}(S)$ the probability that a job finding $j$ other jobs inside the system
will eventually get served. Hence, the probability of abandonment of a job
finding $j$ other jobs ahead of it is simply $1 - P_{j}(S)$. Therefore, we can
write

\begin{equation}
\label{eq:p_j_Ab}
P_{j}(Ab) = \left\{ 
\begin{array}{ll} 
0 & \textrm{if $j < n$}\\
\frac{\displaystyle{(j+1-n)\theta}}{\displaystyle{n\mu + (j+1-n)\theta}}  &
\textrm{if $j \ge n$} 
\end{array} \right. \mbox{.}
\end{equation}

Using Equation~\eqref{eq:p_j_Ab} and the PASTA property (Poisson arrivals see
time averages, see~\cite{mitrani:1998} for more details), $P(Ab)$ yields to

\begin{equation}
\label{eq:P_Ab}
P(Ab) = \sum_{j=0}^{\infty} p_{j}P_{j-n}(Ab)
\mbox{.}
\end{equation}

Computing Equation~\eqref{eq:P_Ab} from its right-hand side is challenging.
However, using Bayes' formula~\cite{ross:2000} we obtain

\begin{equation}
\label{eq:P_Ab1}
P(Ab) = P(W > 0) P(Ab | W > 0) \mbox{,}
\end{equation}

\noindent where $P(W>0)$ is the delay probability, while $P(Ab | W > 0)$ is the
conditional abandonment probability.

In order to compute the above probabilities it is useful to express the balance
equations in terms of $p_{n}$, {\it i.e.}, the
probability that all servers are busy and the queue is empty. 
Using Equations~\eqref{eq:bal_eq} and~\eqref{eq_p_0_1}, $p_{n}$ yields to

\begin{equation}
\label{eq:p_n}
p_{n} = \displaystyle{\frac{B(n,\rho)}{1 + B(n,\rho) \left[\displaystyle{g
\left(\frac{n\mu}{\theta}, \frac{\lambda}{\theta} \right)} -1 \right]}} \mbox{.}
\end{equation}

Hence, Equation~\eqref{eq:bal_eq} can now be written as 

\begin{equation}
\label{eq:bal_eq_1}
p_{j} = \left\{ 
\begin{array}{ll} 
\displaystyle{\frac{n!}{j! \rho^{n-j}}} p_{n} & \textrm{if $j \le n$}\\
\frac{\displaystyle{(\lambda / \theta)^{j-n}}} {\displaystyle{ \prod_{k=1}^{j-n}
\left(\frac{n\mu}{\theta} + k\right)}} p_{n} &
\textrm{if $j > n$} 
\end{array} \right. \mbox{.}
\end{equation}

By means of Equation~\eqref{eq:bal_eq_1} and the PASTA
property, it is now possible to estimate the delay probability

\begin{equation}
\label{eq:PW_pasta}
P(W > 0) = \sum_{j=n}^{\infty} p_{j} = p_{n} + \sum_{j=n+1}^{\infty}
p_{j}\mbox{.}
\end{equation}

Thus, using Equations~\eqref{eq_A}, \eqref{eq:bal_eq_1} and~\eqref{eq:PW_pasta}
we obtain

\begin{equation}
\label{eq:PW}
P(W > 0) = 
g \left(\frac{n\mu}{\theta},\frac{\lambda}{\theta}\right) p_{n} \mbox{,}
\end{equation}

\noindent while, after some manipulations, $P(Ab|W>0)$ yields to

\begin{equation}
\label{eq:cond_ab1}
P(Ab|W>0) = \frac{1}{\displaystyle{\rho g \left(\frac{n\mu}{\theta},
\frac{\lambda}{\theta} \right)}} + 1 - \frac{1}{\rho} \mbox{.}
\end{equation}

{\bfseries N.B.} As the size of the server farms grows, the system
achieves economies of scale that make it more robust against traffic variability. Hence, while violating the
Markovian assumptions about the arrival, patience and service processes affects
the average queue length, it does not substantially change the abandonment
rate~\cite{whitt:2004}.

Having computed the stationary distribution of jobs present and the
corresponding probability of abandonment, we can now compute the average number
of requests served per unit time. That value can be expressed as

\begin{equation}
\label{eq:T}
T = min(n\mu, \lambda [1 - P(Ab)]) \mbox{.}
\end{equation}

{\bf N.B.} The above expression successfully deals with two
special cases, {\it i.e.}, $\theta = 0$, and thus the system would behave
as an $M/M/n$ queue, and $\rho > n$.






\section{Power Usage Estimation}
\label{sec:power}

A number of factors affect the amount of energy consumed by
a server. 
However, the change in the power consumption of a server is mainly due to 
changes in the CPU utilization. 
In order to establish and quantify this relation we have measured the amount of
energy drawn by a server in the presence of an increasing load. 
As for an application we have used 
Wordpress\footnote{\url{http://wordpress.org/}.}, a popular open source
application implementing a blog and running on top of the LAMP\footnote{Linux,
Apache, MySQL and PHP.} stack.
The server had two Xeon Dual Core CPUs (2.8 Ghz) equipped with 
2 Gb of RAM, 7200 RPM hard drive and 1 Gbps network card, while user demand was
generated by Tsung\footnote{\url{http://tsung.erlang-projects.org/}.}. 
The workload consisted of clients arriving over time, with each client simulating a typical behavior of a blog reader, such as checking the front page, navigating through the blog, searching for entries containing certain keywords, etc. The load was increased by reducing the interarrival intervals, which were generated according to an independent Poisson process.
\begin{figure}[ht!]
\centering
\includegraphics[width=0.48\textwidth]{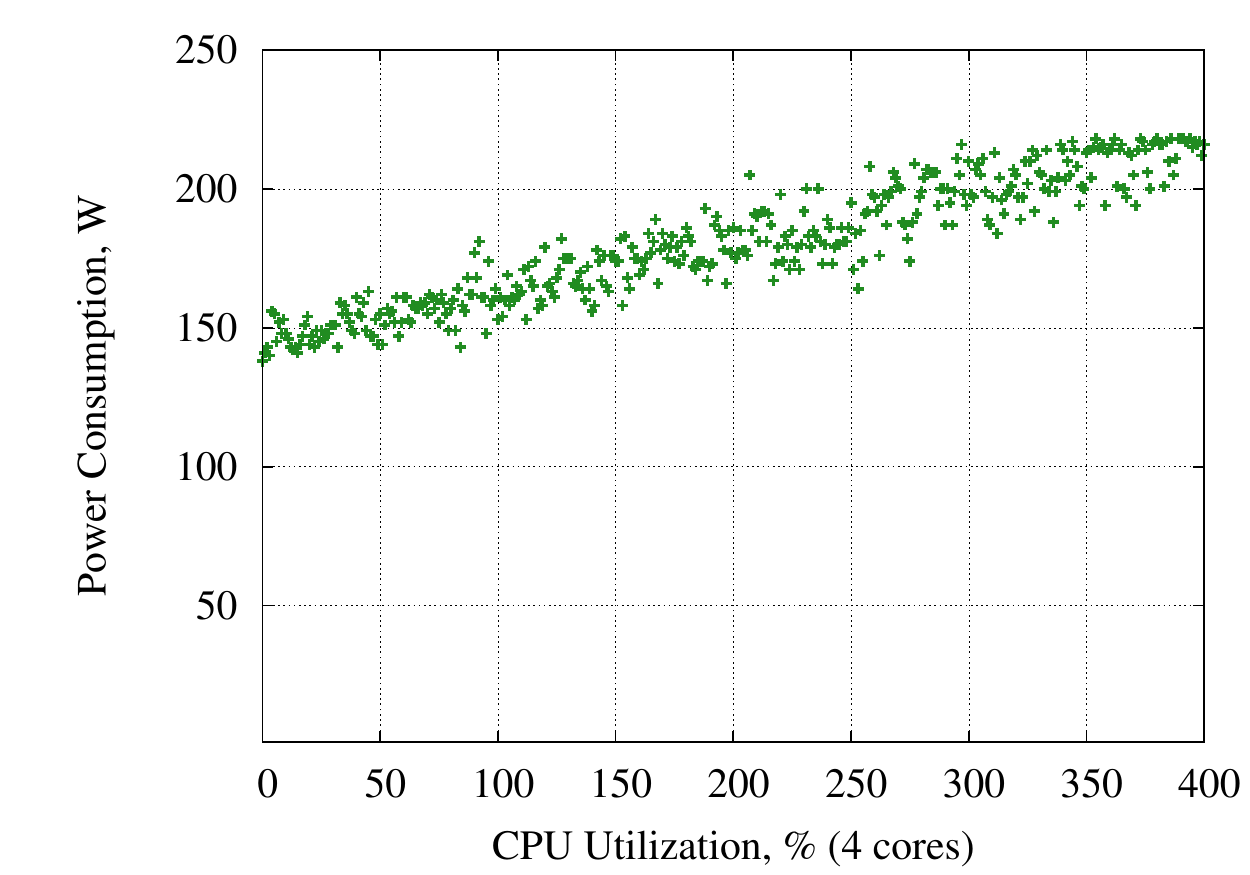}
\caption{Measured energy consumption.}
\label{fig:energy_consumption}
\end{figure}
The measurements reported in Figure \ref{fig:energy_consumption} show a
linear dependency between the CPU utilization and power consumption.
The experiment also confirms that idle servers consume a substantial amount of
energy (140 Watts in our case).
Hence, the average power consumed by a data center per
unit time, $P$, can be estimated as

\begin{equation}
\label{eq:power_linear}
P = n e_{1} + \tau(e_{2} - e_{1}) 
\mbox{,}
\end{equation}

\noindent where $e_{1}$ is the energy consumed per unit time by idle
servers, $e_{2}$ is the energy drawn by each busy server, and $\tau$ is the
occupancy of the system ($\tau \leq n$)

\begin{equation}
\tau = \left \lceil \frac{T}{\mu} \right \rceil \mbox{.}
\end{equation}


\section{Allocation Policy}
\label{sec:policies}

Consider now an allocation decision epoch. The state of the subsystem at
that instant is defined by the number of servers which are not powered down
and by the potential offered load.
If the allocation does not change, the expected revenue for the next
configuration interval is simply $r(n)$. The alternative is to power on/off
some servers. Denote by $n^{\prime}$ the new number
of servers allocated to the queue after reallocation:

Case 1: $n^{\prime} > n$. Such a decision would increase
the system throughput, thus increasing the potential revenue, but it would also
increase the amount of energy consumed by the server farm.

Case 2: $n^{\prime} < n$. Such a decision would decrease the expenditure for
electricity, as $(n - n^{\prime})$ servers would be switched off. However, it would
also decrease the overall throughput, thus increasing the probability of jobs
abandonment.

Powering servers on/off is not instantaneous, but takes on average $k$ units
time. Hence, at every state change there is some waste of time and money
because servers changing their state can not serve user demand, but do consume electricity. 
Also, system's reliability is affected by state changes, as
hardware components tend to degrade faster with frequent power on/off cycles
than with continuous operation. For example, hard disks have an average lifetime
of 40/50,000 on/off cycles~\cite{greenawalt:1994}. Therefore, each state change
involves the following cost

\begin{equation}
\label{eq:q}
Q =  \frac{|\Delta n|}{t}
(\sum_{i=1}^{l} d_{i} + kre_{max}) \mbox{,}
\end{equation}

\noindent where $t$ is the length of the observation windows,
$|\Delta n|$ is the number of servers that are switched on/off, $e_{max}$ is
the power consumed per unit time during state changes, $k$ is the average time
required to switch a server on/off, $d_i$ is the cost for a hardware
component's state change, and $l$ is the number of components.

{\bf N.B.} One can easily relax the assumption that powering a server up and
down takes the same amount of time and consume the same amount of energy. 

The expected {\it change} in revenue resulting from a decision to
change the number of running servers can be expressed as

\begin{equation}
\label{eq:deltaR}
\Delta r(n^{\prime}, n) = r(n^{\prime}) - r(n) - Q \mbox{.}
\end{equation}


When $R$ is computed for different values of $n$, it becomes clear that the revenue is
a unimodal function of $n$, {\it i.e.}, it has a single maximum.
That observation implies that one can search for the optimal number of servers
to run by evaluating $R$ for consecutive values of $n$, stopping either when
$R$ starts decreasing or, if that does not happen, when the increase becomes
smaller than some value $\epsilon$. This can be justified arguing that $R$ is a
concave function with respect to $n$. Intuitively, the economic benefits of
powering more servers on become less and less significant as $n$ increases,
while the loss of potential revenues gets bigger and bigger as $n$ decreases.
Such a behavior is an indication of concavity.
Thus, a fast algorithm of the {\it binary search} variety suggested by the
above observations works as follows:

\begin{enumerate}
\item Set with $n_{l} = 0$ and $n_{u} = S$;
\item Set $n^{\prime} = \lceil \lambda / \mu \rceil$. If  $n^{\prime} = 0$, set  $n^{\prime} = 1$. Similarly, if  $n^{\prime} = S$, set  $n^{\prime} = S -1$.
\item While $n_{l} < n_{u}$ \label{alg:while}
\begin{enumerate}
	\item Calculate $\Delta r(n^{\prime} - 1, n)$, $\Delta r(n^{\prime}, n)$ and	$\Delta r(n^{\prime} + 1, n)$.
	\item If $\Delta r(n^{\prime} - 1, n) \le \Delta r(n^{\prime}, n) \ge \Delta r(n^{\prime} + 1, n)$, then $n^{\prime}$ is the best solution. Hence, go to~\ref{alg:comparison}.
	\item If  $\Delta r(n^{\prime} - 1, n) \le \Delta r(n^{\prime}, n) \le \Delta r(n^{\prime} + 1, n)$, then the optimal $n$ is in the interval $n^{\prime}, \ldots, n_{u}$. Hence, set $n_{l} = n^{\prime} + 1$ and go to~\ref{alg:newn}.
	\item We have  $\Delta r(n^{\prime} - 1, n) \ge \Delta r(n^{\prime}, n) \ge \Delta r(n^{\prime} + 1, n)$. Hence, the search has to be carried out in the interval $n_{l}, \ldots, n^{\prime}$, so set $n_{u} = n^{\prime} - 1$ and go to~\ref{alg:newn}.
\item Set $n^{\prime} = \lceil n_{l} + (n_{u} - n_{l}) / 2 \rceil$.\label{alg:newn}
\end{enumerate}
\item If $\Delta r(n^{\prime}, n) > 0$, then set $n = n^{\prime}$. Otherwise, leave the allocation as it is \label{alg:comparison}.
\end{enumerate}

Since at every iteration the state space is reduced by a factor of two, 
$log(S)$ iterations are required in order to find the `best' $n$, {\it
i.e.}, the number of servers that maximizes the profit. We have put quotation
marks around the word `best' because such choice might be slightly sub-optimal
when the exponential assumptions are violated and $S$ is small (see the remarks
at the end of Section~\ref{sec:analysis}). This policy will be referred to as the `Adaptive' allocation policy.

Finally, since the state change is not instantaneous, the current window ends
immediately if some of the servers are to be switched off ({\it e.g.}, if
$n^{\prime} < n$) or after all the servers are powered up if $n^{\prime} > n$.

\section{Results}
\label{sec:experiments}
 
Several experiments were carried out, with the aim of evaluating the effects of
proposed scheme on the maximum achievable revenues.
We assume the server farm has a Power Usage Effectiveness (PUE), the main
metric used to evaluate the efficiency of data centers, of 1.7. That value is
computed as the ratio between the total facility power and the IT equipment
power. Also, we take indirect costs into account. These include the cost
for capital as well as the amortization of the equipment such as servers, power
generators or transformers, and account for twice the cost of consumed
electricity.
Finally, in order to reduce the number of variables, when not specified otherwise, the following features were held fixed:
\begin{itemize}
  \item	250 servers, configured as described in Section~\ref{sec:power}, {\it
  i.e.}, $S = 1,000$.
  \item The power consumption of each four core server machine ranges
	between 140 and 220 W, see Figure~\ref{fig:energy_consumption}. In other 
	words, each core from now on server consumes 
	between 35 and 55 W.
	Since the server farm has a PUE factor of 1.7, the minimum and maximum power
	consumption are approximately $e_{1}$ = 59 and $e_{2}$ = 94 W per
	server.
  \item The cost for electricity, $r$, is 0.1 \$ per
  kWh\footnote{http://www.neo.ne.gov/statshtml/115.htm.}.
  \item The average job size, $1 / \mu$, is 0.1 seconds.
  \item Jobs are not completely CPU bound. Instead, when a server is busy, the
average CPU utilization is 70\%. In other words, busy servers draw~69.58
Wh, and thus each job costs, for electricity, $2 \times 10^{-7}$\$ on
average.
	\item Each successfully completed job generates, on average, $6.2 \times 10^{-6}$\$.
\end{itemize}

It is worth noting that while the size of the server farm might look
small, the application logic of the 10th busiest web site in the world, Wikipedia, is hosted on 350 servers having 856 cores, spread across three data
centers~\cite{bergsma:2007}.

\subsection{Stationary Traffic}

\begin{figure}[h!]
\centering
\includegraphics[width=0.48\textwidth]{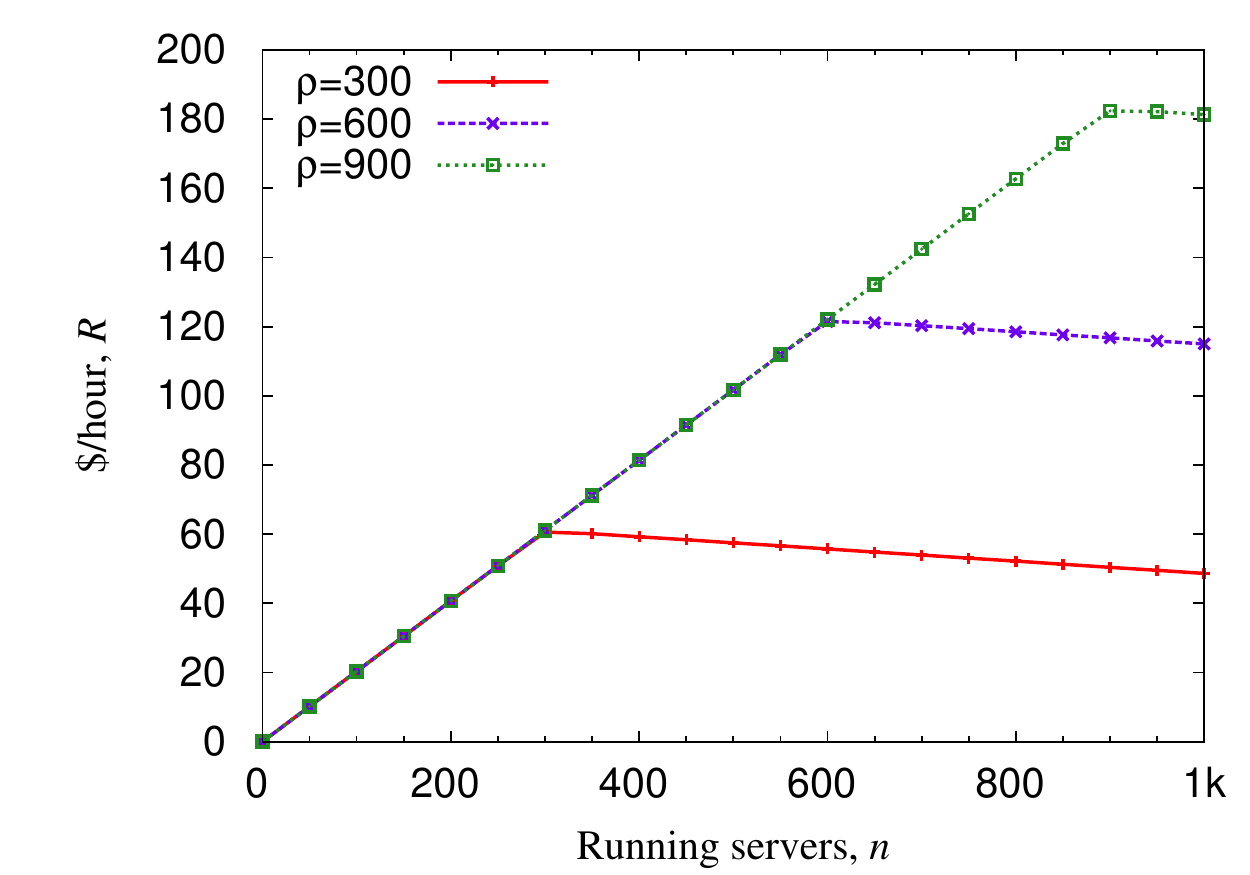}
\caption{Revenue as a function of the running servers.}
\label{fig:exp1}
\end{figure}

The first experiment, see Figure~\ref{fig:exp1}, is purely numerical. 
It shows how the number of running servers affects the average earned revenue
under different loading conditions.
The potential offered load is increased from 30\% to 90\% by
increasing the rate at which new jobs enter the system, from 3,000 to 9,000
jobs per second.

Figure~\ref{fig:exp1} shows that $(i)$ in each case there is an optimal number
of servers that should be switched on; $(ii)$ the heavier is the load, the
higher is the optimal number of servers to run, but higher is also the maximum achievable revenue;
	$(iii)$ when $n > n_{opt}$, the system under-performs because the cost of
  	running idle servers erodes revenues, while when $n < n_{opt}$, the system
  	under-performs because it misses potential revenues.

Next, we evaluate the effectiveness of the dynamic allocation scheme via
computer simulation.
For comparison reasons, two versions of the `Static' policy, a policy which runs
always the same amount of servers, are also displayed. One runs $n
= S/2 = 500$ servers, while the other $n = S = 1,000$. 
We vary the load between 10\% and 110\% ({\it i.e.}, the system would be over-saturated without job abandonment) by varying the arrival rate, {\it i.e.},  $\lambda =
1,000, \ldots, 11,000$ jobs/second. 
Each point in the figure represents one run lasting 16.5 hours, while servers
are reallocated every hour, {\it e.g.}, the policy described in
Section~\ref{sec:policies} is invoked every hour. During each run, approximately between 119 (low load)
and 653 (high load) million jobs enter the system, while samples of achieved revenues are collected every 1.5 hours and are used at the end of each run to compute the corresponding 95\% confidence interval, which is calculated using the Student's t-distribution.

\begin{figure}[h!]
\centering
\includegraphics[width=0.48\textwidth]{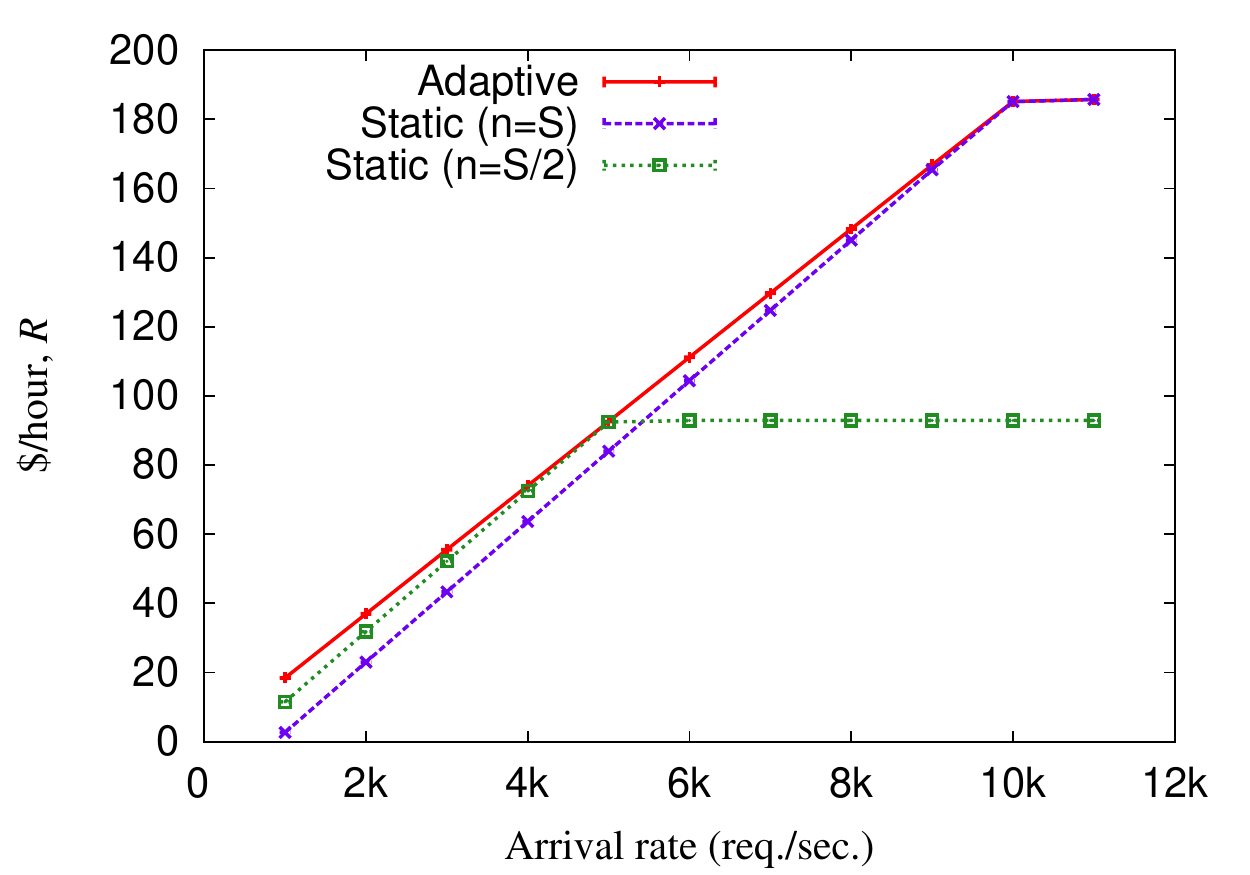}
\caption{Observed revenues for different policies.}
\label{fig:markovian_r}
\end{figure}

\begin{figure}[ht!]
\centering
\includegraphics[width=0.48\textwidth]{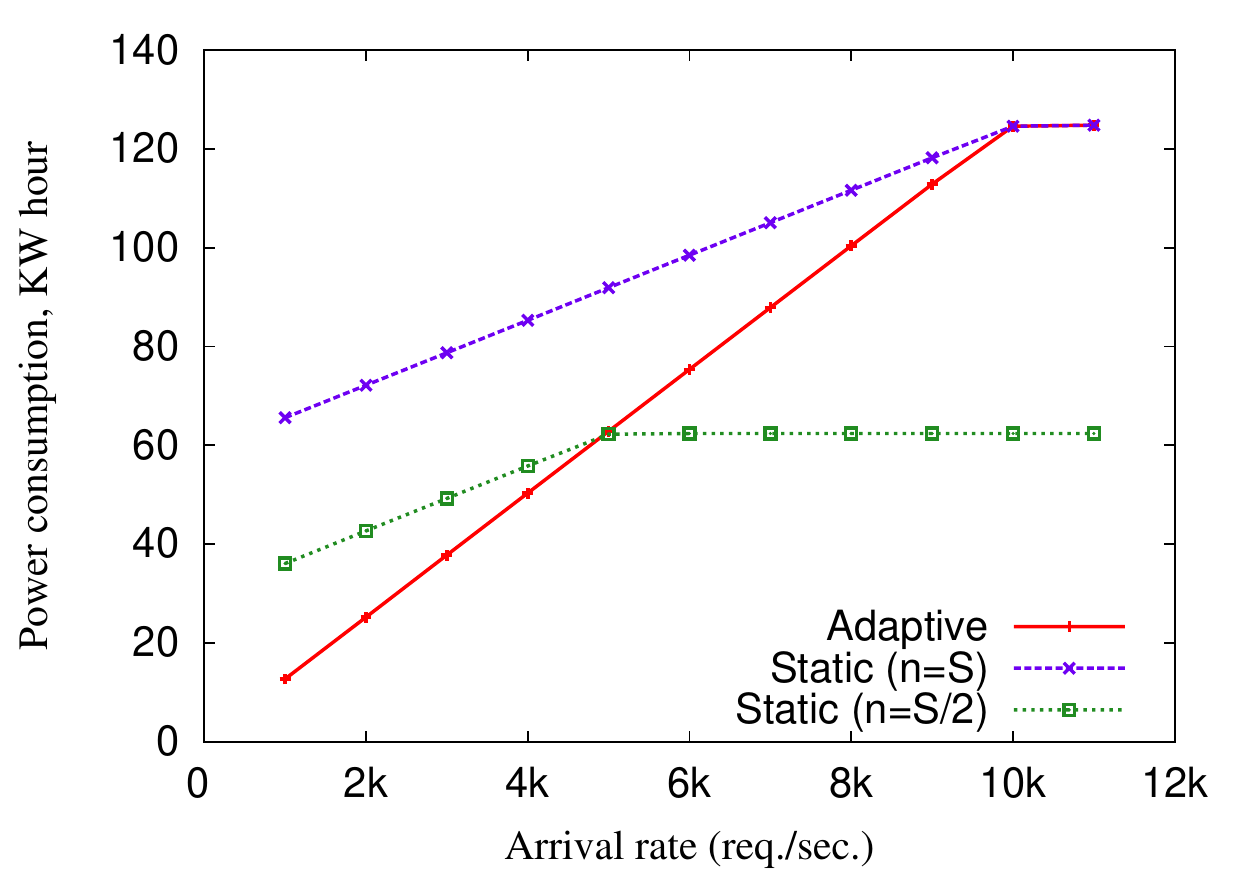}
\caption{Observed energy consumption for different policies.}
\label{fig:energy}
\end{figure}

The most notable feature of the graph plotted in Figure~\ref{fig:markovian_r}
is that the `Static' policies do not perform well 
under light load (because of the servers running idle), while the one
with parameter $n=S/2$ can not cope with high traffic, thus missing income opportunities. 
On the other hand, the `Adaptive' heuristic produces revenues that grow with the offered load.
%
%
%
%
Next, we report the average power consumption. Figure~\ref{fig:energy} shows
that the `Adaptive' heuristic runs servers only when necessary, thus reducing
its carbon footprint as well as the provider's electricity bill.

In the next experiment we evaluate the effects of increased variability on the performance of the `Adaptive' heuristic by departing from the assumption that the interarrival and patience times are exponentially distributed. Now those parameters are Log-Normally distributed, with the squared coefficient of variation, {\it i.e.}, the variance divided by the square of the mean, of 2 (interarrival intervals on YouTube have a squared coefficient of variation of  1.7~\cite{phillipa:2007}) and 5, respectively.
These changes increase the variances of the interarrival and patience times while preserving the averages, thus making the system less predictable and allocation decisions more challenging.

\begin{table}[ht!]
\begin{tabular}{rrrrrrr}
 & \multicolumn{3}{r}{$scv(\lambda)=scv(\theta)=1$} & \multicolumn{3}{r}{$scv(\lambda)=2, scv(\theta)=5$}\\
\cline{2-4}
\cline{5-7}
S & R & L & \% Ab & R & L & \% Ab\\
\hline
10 & 1.45 & 4.636	& 1.136 & 1.44 & 6.216 & 1.398\\
20 & 2.91 & 4.602	& 0.572 & 2.91 & 7.040 & 0.614\\
50 & 7.35 & 11.034 & 0.547 & 7.35 & 17.374 & 0.514\\
100 & 14.75 & 14.371	& 0.359 & 14.76 & 24.007 & 0.268\\
500 & 74.06 & 28.467	& 0.142 & 74.11 & 53.349 & 0.055\\
1000 & 148.26 & 40.651 & 0.101 & 148.38 & 82.370	& 0.033\\
\hline
\end{tabular}
\caption{Performance of the `Adaptive' heuristic for different parameters.}
\label{tab:mark_vs_non_mark}
\end{table}

In Table~\ref{tab:mark_vs_non_mark} we compare the achieved revenue (R), average queue length (L) and the percentage of jobs abandoning the system under the `Adaptive' policy for different values of $S$. The system is always loaded at $80\%$ by scaling the arrival rate in proportion to $S$.
From the experiment we observe that while $L$ depends on the variability of the traffic parameters, the achieved revenue and the percentage of jobs leaving do not. Moreover, the system becomes less and less sensible to the distribution of the interarrival intervals and patience times as $S$ increases.
This can be explained by observing that large systems can achieve economies of scale which are simply not possible when the number of servers is small.
In particular, the behavior of large server farms under heavy load 
differs from that of Kingman's Law ({\it i.e.}, delays/job losses are
very common under heavy load) in that service quality is carefully balanced with server efficiency.

\subsection{Sensitivity Analysis}

The previous experiments investigated how server allocation decisions and 
other parameters can affect the revenue. However, in  real world scenarios
the chances that service providers would have to deal with stationary traffic
are extremely rare. In fact, studies of the Wikipedia traces show that
throughout the day the incoming traffic can change by as much as
70\%~\cite{urdaneta:2009}. A logical solution to this problem would be reconfiguring the servers pool (by
switching servers on/off) according the changes in the load, as we propose in Section~\ref{sec:policies}. However, service
providers need to estimate the arrival rate for the next configuration interval. 
Such a prediction might be hard to make if the load is non stationary, and even good forecasting tools might 
produce results which sometimes differ from the observed values.
Forecasting tools range from simply using the last observed value of
$\lambda$ to very complex algorithms requiring significant periods of time for
training and considering seasonal and trend components. Instead of benchmarking
the proposed approach with various forecasting tools, which would lead to an explosion of the experimental space, in the following we
investigate its sensitivity in respect to an error in load prediction. This also
can help in choosing the right prediction mechanism in real-world scenarios, as
the deployment of certain forecasters may require long training, while simpler
ones might exhibit slightly worse prediction quality, while having little or no
effect on the ultimate result.

In order to see how various error rates affect the revenue, we introduce an
Oracle forecaster which tells the exact arrival rate for the next configuration
interval. Then we introduce an error, first 5\%, then 10\% and finally 20\%.
Please note that the forecasting values are distributed according to a Laplace
distribution with the mean value equal to the actual $\lambda$. Thus, for example
5\% represents the mean of the absolute differences between actual and predicted
values of $\lambda$. We have chosen a Laplace distribution because we have
observed such a behavior when trying to predict the Wikipedia workload using the
double exponential smoothing method. Unfortunately, due to the lack of space we cannot present this data. Also, it is important to note that by combining Oracle with the binary search from section \ref{sec:policies} we obtain the optimal solution in terms of profit
maximization.

In the next set of experiments we contrast five different cases: static policy
(all servers are switched on), Adaptive heuristic (see
Section~\ref{sec:policies})  with Oracle forecasting, and Adaptive  heuristic
with Oracle forecasting making a systematic error of 5\%,10\% and 20\%. 
In the simulation, a workload which represents a scaled version of the
Wikipedia traces is employed. The arrival rate ranges between 2,688 and 
5,729 jobs/second, while each simulation run lasts 240 hours ({\it e.g.}, 10 days).


\begin{table}[ht!]
\centering
\caption{Percentage of lost jobs with respect to error in the prediction}
\label{tbl:cum_jobs_lost}
\begin{tabular}{l|r|r|r|r}
Error & 0\%	& 5\% &	10\% & 20\% \\
\hline
Lost jobs, \% & 0.01	& 2.19 & 4.46	& 10.54 \\

\end{tabular}
\end{table}

\begin{figure}[ht!]
\centering
\includegraphics[width=0.48\textwidth]{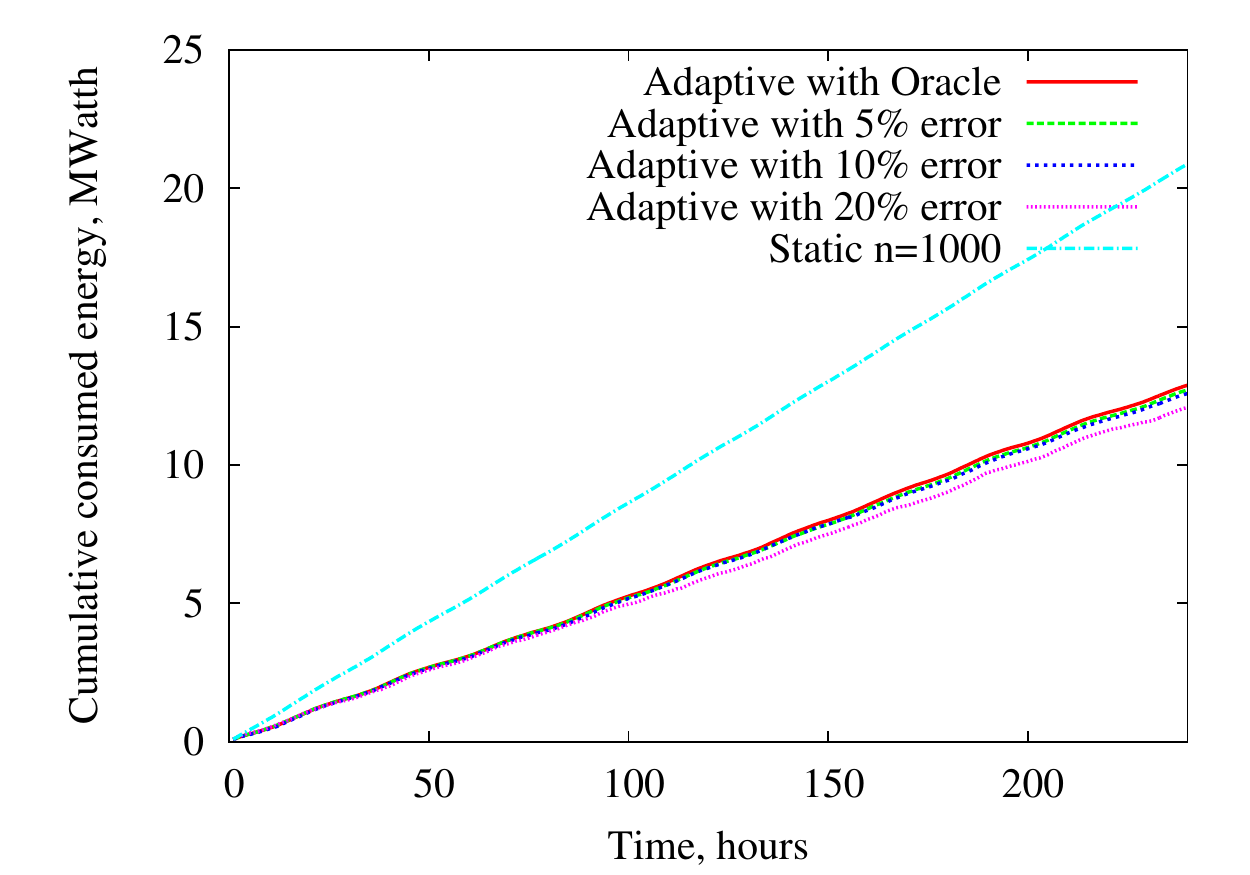}
\caption{Cumulative energy consumption.}
\label{fig:cum_energy}
\end{figure}

\begin{figure}[ht!]
\centering
\includegraphics[width=0.48\textwidth]{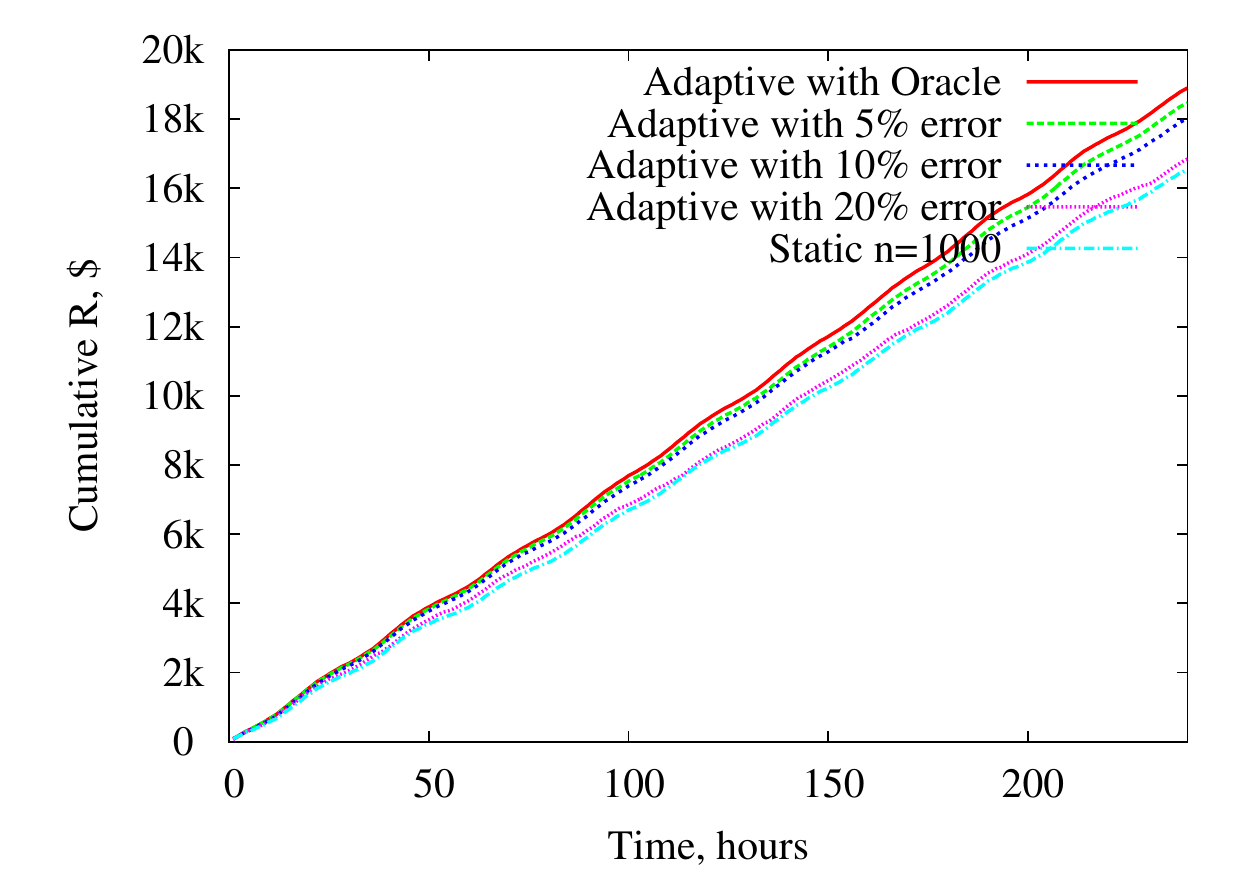}
\caption{Cumulative revenue.}
\label{fig:cum_rev}
\end{figure}

As shown in Table~\ref{tbl:cum_jobs_lost}, the error in the
forecasting has a significant impact on the number of lost jobs. 
The Oracle behaves almost as good as the static policy which massively
over-provisions the system, which in its turn negatively reflects on the amount
of consumed energy, see Figure~\ref{fig:cum_energy}, and consequently the
revenue, see Figure~\ref{fig:cum_rev}.

At the same time 5\% error in the prediction affects the number of lost
jobs without significant impact on the other metrics. An increase of the error
from 5\% to 10\% markedly reflects on the number of lost jobs, while the revenue
values still stay very close. A jump of the error from 10\% to 20\% not only
increases the number of lost jobs more than twice due to frequent mistakes
causing under-provisioning and thus adversely affecting the revenue. Despite the
obvious advantage in the power consumption, using forecasting with 20\% error is
unacceptable due to its low revenue and high percentage of the lost jobs.
From the above experiment we can conclude that 
5\% error in forecasting does not have significant impact on the revenue and
 energy consumption, while a 10\% error still exhibits results which are markedly superior to the Static
  allocation policy.

\section{Conclusions}
\label{sec:conclusions}

In this paper we have introduced and evaluated an easily implementable policy for
dynamically adaptable Internet services. Under some simplifying assumptions, the
numerical algorithm we propose can find the best trade off between consumed power
and delivered service quality. We have demonstrated that the number of running
servers can have a significant effect on the revenue earned by the provider.
The experiments we have conducted show that our approach works well under
different traffic conditions, and that our policy is not very sensitive to
errors in parameters estimation.


Possible directions for future research include taking into account the trade
offs between the number of running servers, the frequency of the CPUs and the
maximum achievable performance, as well as fault tolerance issues.

\section*{Acknowledgements}

The authors would like to thank the European Commission (Marie Curie
Action, contract number FP6-042467), the EU Cost Action IC0804, and the EUREKA
Project 4989 (SITIO).


\bibliographystyle{IEEEtran}
\bibliography{green-computing}

\end{document}